\documentclass{elsart}
\usepackage{epsfig}
\usepackage{amsmath}
\usepackage{amsfonts}
\usepackage{color}
\usepackage{graphicx,slashbox}
\usepackage{ulem}

\newcommand{\JMcomm}[1]{{\color{black} #1}}
\newcommand{\PGRcomm}[1]{{\color{black} #1}}
\newcommand{\MDcomm}[1]{{\color{black} #1}}

\begin{document}

\begin{frontmatter}

\title{Improved Slater approximation to SIC-OEP}

\author{J. Messud$^a$}
\author{, P.~M.~Dinh\corauthref{cor}$^a$}
\author{, P.-G.~Reinhard$^b$, and E.~Suraud$^a$}

\corauth[cor]{Corresponding author\\{\it Email-address}~:
  dinh@irsamc.ups-tlse.fr} 
\address{$^a$Laboratoire de Physique Th\'eorique, Universit\'e Paul
  Sabatier, CNRS, F-31062 Toulouse C\'edex, France}
\address{$^b$Institut f{\"u}r Theoretische Physik, Universit{\"a}t
  Erlangen, D-91058 Erlangen, Germany}

\begin{abstract}
We propose a simplification of the Optimized Effective Potential (OEP)
method applied to the Self Interaction Correction (SIC) on the Local
Density Approximation (LDA) in Density Functional Theory (DFT). The
new scheme fulfills crucial formal key properties. It turns out to be
simple and accurate.  We apply the new method to a schematic model for
a dimer molecule and to the C atom.  We discuss observables which are
especially sensitive to details of the SIC.
\end{abstract}

\end{frontmatter}

\section{Introduction}

Density-functional theory (DFT) is a major tool for the description of
electronic systems as, e.g., atoms, molecules, solids, or chemical
reactions \cite{Koh99aR}.
It simplifies the involved many-electron problem and thus allows to
compute rather complex systems, for a general overview see
\cite{Par89B,Dre90B}. Most applications employ energy-density
functionals based on the Local Density Approximation (LDA), see
e.g. \cite{Jon89aR}, or its extension to the generalized gradient
approximation (GGA) \cite{Per96a}. In spite of their success, these
approaches have still deficiencies. In particular, the
self-interaction error spoils single-particle properties as, e.g., the
Ionization Potential (IP) or the band gap in solids
\cite{Hyb86a,Nie00aR}. Another critical detail where LDA and GGA fail
is the polarizability in chain molecules \cite{Gis99a,Kue04a}.
A way out of the self-interaction problem is to use exact
exchange. That, however, is an order of magnitude more expensive and
it causes problems with a reliable description of the remaining
correlation effects \cite{Kue07aR}. An intuitive and efficient
solution, still close to the spirit of DFT, is to augment LDA by a
Self-Interaction Correction (SIC) \cite{Per79a,Per81a}. That, on the
other hand, produces a state-dependent mean-field Hamiltonian which
requires extra measures to maintain orthogonality of the single
particle basis \cite{Per81a,Har83a,Sva90a,Goe97a}. The desirable goal
remains to have a common local mean-field potential $V(\mathbf{r})$.
That goal is reached through the optimized effective potential (OEP)
method, see \cite{Sha53a,Tal76} for early proposals, and
\cite{Kue07aR} for an extensive recent review. There exist OEP
approaches based on exact exchange but these are still rather
involved, as exchange is. In that case, OEP serves mainly to produce a
well defined single-particle spectrum in occupied space as well as in
unoccupied space \cite{Goe99a,Del01a}. But OEP is also a useful means
to overcome the difficulties of SIC, see
e.g. \cite{OEP1,OEP2,TDOEP}. Indeed it is found that OEP manages to
maintain crucial features of the underlying SIC (or exact exchange)
as, e.g., the localization of states or the derivative discontinuity
\cite{Per83a,Kri90a,Mun05a}.  The involved SIC-OEP equations are thus
often simplified. Popular is, e.g., the Krieger-Li-Iafrate (KLI)
approach \cite{OEP1,OEP2} and, in a further step of simplification,
the Slater approximation \cite{Sha53a}. However, KLI and Slater
approximation can easily miss crucial features of SIC as, e.g., the
localization of states and the performance with respect to
polarizability \cite{Gis99a,Kue04a}.

It is the aim of this paper to propose a scheme for SIC-OEP which
allows to deal with the Slater approximation to OEP while
maintaining key features of the full SIC scheme. The new move is to
allow for a double set of occupied single-particle states. The two
sets are connected by a unitary transformation such that both sets
build the same total density.
The first of the sets consists of the solutions of the SIC-OEP
equations and is thus diagonal in energy. The second set is used to
compute the total SIC energy.  The distinction between the two
sets allows the second set to develop spatially localized states which
is energetically advantageous for the SIC energy functional.  The
localization of the second set validates the Slater approximation to
OEP. We call the new scheme ``generalized Slater approximation''.  We
present it after briefly reviewing SIC-OEP, SIC as such, and a variant
of SIC which already employs the double set of single-particle
states. The method is applied to a schematic model for a dimer
molecule and to the C atom. We will compare exact Hartree-Fock with
all relevant approximations (LDA, SIC, Slater and generalized Slater).
In all the following, we employ, in fact, the local spin-density
approximation but denote it with the simpler term LDA (rather than the
sometimes used more complete abbreviation LSDA).

\section{SIC without approximations}

The starting point for all following considerations is the
SIC energy functional for electrons, which reads
\begin{eqnarray}
  E_\mathrm{SIC}
  =
  E_\mathrm{kin}
  \!+\!
  E_\mathrm{ion}
  \!+\!
  E_\mathrm{LDA}[\rho]
  \!-\!
  \sum_{\beta=1}^{N} E_\mathrm{LDA}[\rho_{\beta}]
\label{eq:fsicen}
\end{eqnarray}
where $E_\mathrm{LDA}[\rho]$ is a standard energy-density functional
in LDA complementing the kinetic energy $E_\mathrm{kin}$ and the
interaction energy with the ionic background $E_\mathrm{ion}$. The
last term therein is the SIC term. The densities are defined from the
single-particle states $\psi_\beta$, i.e. $\rho=\sum_{\beta}
\rho_{\beta}$ with $\rho_\beta=|\psi_{\beta}|^2$.  All summations run
over occupied states only.

Two features of the SIC energy (\ref{eq:fsicen}) are to be noted.
First, it is not invariant under unitary transformations amongst the
occupied states because the SIC term is sensitive to each
single-particle state separately.  Second, the SIC induces a tendency
to localized single-particle states. Given a total density $\rho$, the
distribution over several $\rho_\beta$ produces maximal Coulomb energy
if these $\rho_\beta$ are localized.  Such a localization is found in
many studies of SIC~\cite{Ped84} or SIC-OEP \cite{Kue07aR}. This
feature will play a crucial role in the following considerations.

The mean-field equations are derived from the total energy
(\ref{eq:fsicen}) by variation with respect to the single-particle
states $\psi_{\alpha}^*$.  In the general case, the mean-field
Hamiltonian becomes 
\begin{subequations}
\label{eq:SICmfham}
\begin{eqnarray}
  \hat{h}_\alpha
  &=&
  \hat{h}^\mathrm{(LDA)}
  -
  U_\alpha(\mathbf{r})
  \quad,
\\
  U_\alpha(\mathbf{r})
  &=&
  \frac{\delta E_\mathrm{LDA}[\rho_\alpha]}
       {\delta \rho_\alpha(\mathbf{r})}
  =
  U_{\rm LDA} \left[ |\psi_\alpha|^2 \right]
  \quad,
\label{eq:SICmf}
\\
  \hat{h}^\mathrm{(LDA)}
  &=&
  \frac{\hat{p}^2}{2m}
  +
  \frac{\delta E_\mathrm{LDA}[\rho]}
       {\delta \rho(\mathbf{r})}
  \quad,
\end{eqnarray}
\end{subequations}
where the first term is the standard LDA mean field
and the second term, state-dependent, stems from the SIC term in the
energy  (\ref{eq:fsicen}).

\subsection{SIC with single set}
\label{eq:SICsingle}

The SIC mean-field Hamiltonian (\ref{eq:SICmfham}) is state-dependent.
A direct solution of the Schr\"o\-din\-ger equation with a
state-dependent Hamiltonian can lead to non-orthogonal single-particle
states. This, however, violates basic requirements of DFT. One has to
enforce orthonormality by a constraint, thus minimizing
$
E_\mathrm{SIC} - 
\sum_{\alpha\beta}\lambda_{\alpha\beta}(\psi_\alpha|\psi_\beta)
$
where $\lambda_{\alpha\beta}$ is a matrix of Lagrangian multipliers,
which is non-diagonal in general.
By straightforward variational techniques, we then obtain the stationary
SIC equations~:
\begin{subequations}
\label{eq:statsic1}
\begin{eqnarray}
  \hat{h}_\mathrm{SIC}|\psi_\alpha)
  &=&
  \sum_{\beta} |\psi_{\beta})
  \lambda_{\beta\alpha} \ ,
\label{eq:statsic}
\\
  \lambda_{\beta\alpha}
  &=&
  (\psi_\beta|\hat{h}_\mathrm{SIC}|\psi_\alpha) \ ,
\label{eq:constrmat}
\\
  0
  &=&
  (\psi_\beta|U_\beta-U_\alpha|\psi_\alpha) \ ,
\label{eq:symcond}
\\
  \hat{h}_\mathrm{SIC}
  &=&
  \hat{h}_\mathrm{LDA}
  -
  \sum_\alpha U_\alpha|\psi_\alpha)(\psi_\alpha| \ ,
\label{eq:hsic}
\end{eqnarray}
\end{subequations}
where we have packed the state-dependence into projector notation in
the SIC mean field (\ref{eq:hsic}).  This yields a formally compact
notation of $\hat{h}_\mathrm{SIC}$.  Note that Eq.~(\ref{eq:statsic})
is not an eigenvalue equation because the matrix of Lagrange
multipliers may possibly become non-diagonal. This equation as such
thus does not provide a valuable starting basis for developing an OEP
scheme, which requires implicitly an eigenvalue equation \cite{Tal76}.
A key part in the SIC equations (\ref{eq:statsic1}) 
is the symmetry condition (\ref{eq:symcond}).  
It emerges as follows~: The Lagrangian matrix is real symmetric,
$\lambda_{\alpha\beta}=\lambda_{\beta\alpha}$, because the
overlap matrix $(\psi_\alpha|\psi_\beta)$ has that feature.
Thus the hermitian conjugate of (\ref{eq:statsic})
becomes 
$(\psi_\beta|\hat{h}_\mathrm{SIC}^\dagger
  =
  \sum_{\alpha}\lambda_{\beta\alpha}(\psi_{\alpha}|\,.
$
We take the difference. The $\lambda_{\beta\alpha}$ cancel out
and we remain with
$
 0
 =
 (\psi_\beta|\hat{h}_\mathrm{SIC}^\dagger
 -
  \hat{h}_\mathrm{SIC}|\psi_\alpha)
$
which then yields the symmetry condition (\ref{eq:symcond}).  
It is a highly non-linear equation requiring an involved solution
strategy. But formally, the symmetry condition simply emerges from
minimizing the total energy in a reduced space of orthonormalized
single-particle orbitals. This minimization principle 
also implies that there always exists a
solution to the symmetry condition, in spite of its apparent complexity.

\subsection{SIC with double basis set}
\label{sec:sicdouble}

The SIC equations (\ref{eq:statsic1}) do not yield immediately
single-particle energies and associated ``energy-diagonal'' states. It is
plausible that the matrix $\lambda_{\beta\alpha}$ contains the
relevant information.  The idea is thus that diagonalization of
$\lambda_{\beta\alpha}$ yields the desired single-particle energies
together with energy-diagonal states $\varphi_i$. These energy-diagonal
states tend naturally to larger spatial width and will not remain as
localized as the SIC basis states $\psi_\alpha$. Both sets have
their virtues but they cannot be identical due to the state-dependence
of the SIC mean field.
This suggests to introduce throughout a second basis set
$\varphi_i$ which is related to the set $\psi_\alpha$ by a
unitary transformation within occupied space, i.e.
\begin{subequations}
\label{eq:statsic2}
\begin{equation}
  \psi_\alpha
  =
  \sum_{i=1}^N \varphi_i \, u_{i\alpha}
  \quad.
\label{eq:unitrans}
\end{equation}
We exploit the freedom of a unitary transformation 
$u_{i\alpha}$ to {choose} a basis which
diagonalizes the constraint matrix $\lambda_{\beta\alpha}$.
This then yields the diagonal SIC equations
\begin{equation}
  \hat{h}_\mathrm{SIC}|\varphi_i)
  =
  \varepsilon_i |\varphi_i)
  \quad.
\label{eq:static-diaq}
\end{equation}
together with the symmetry condition 
\begin{equation}
  u_{i\alpha}\,:\quad
  0
  =
  (\psi_\beta|U_\beta-U_\alpha|\psi_\alpha) \quad,
\label{eq:symcond2}
\end{equation}
\end{subequations}
which now serves to determine the coefficients $u_{i\alpha}$ of the
transformation (\ref{eq:unitrans}) for given $\varphi_i$.  Note that
the symmetry condition (\ref{eq:symcond2}) looks similar to
Eq.~(\ref{eq:symcond}), but is used somewhat differently here, namely
to define the coefficients of the unitary transformation
(\ref{eq:unitrans}). The $\hat{h}_\mathrm{SIC}$ is defined in detail
by the SIC equations (\ref{eq:statsic1}).  The new
Eqs.~(\ref{eq:statsic2}) combine the solution of the former SIC
equations, yielding the $\psi_\alpha$, with the direct evaluation of
the energy-diagonal basis. What expense is concerned, there is no
advantage as compared to the previous scheme, i.e. solving first the
SIC equations and diagonalizing then the constraint matrix
$\lambda_{\beta\alpha}$ in a second step. But it serves from a formal
point of view as a preparatory step for SIC-OEP.  Although
Eq.~(\ref{eq:static-diaq}) is an eigenvalue equation, the
corresponding Hamiltonian $\hat{h}_{\rm SIC}$ is obviously non-local,
see Eq.~(\ref{eq:hsic}), which complicates the numerical
handling. This is why approximations have been searched for since long
in order simplify calculations. 

\section{SIC-OEP}
\label{sec:SIC-OEP}

\subsection{SIC-OEP with double set}
\label{sec:sic_oep_2sets}

In section \ref{sec:sicdouble}, we have shown that the definition of
single-particle energies in connection with SIC naturally leads to the
introduction of a double set of single-particle states connected by a
unitary transformation. We transfer now this generalization to OEP.
The idea beyond SIC-OEP is the following. We start from a 
set ${ {\tilde \varphi_i}}$, being solution of the OEP
equations~:
\begin{equation}
  \left[\hat{h}^\mathrm{(LDA)}({\bf r})-V_0(\mathbf{r})\right] {\tilde
  \varphi_i}({\bf r})
  =
  \varepsilon_i {\tilde \varphi_i}({\bf r})
  \quad,
\label{eq:eigen3}
\end{equation}
where $V_0$ is a local and state-independent potential which needs to
be optimized to minimize the SIC energy, Eq.~(\ref{eq:fsicen}). Formally,
this amounts to parameterize the SIC wavefunctions through $V_0$, to
consider the SIC energy as a functional of $V_0$, and to
perform variation  with respect to  $V_0$.
The $|{\tilde\varphi_i})$ can thus be written as $|\varphi_i^{V_0})$
and these are associated to the corresponding set of
$|{\psi_\alpha^{V_0}})$ connected by the unitary transformation
Eq.~(\ref{eq:unitrans}). For the sake of simplicity, we keep in the
following the notation $|{\tilde \varphi_i})$ for the
$|\varphi_i^{V_0})$ and use correspondingly $|{\tilde\psi_\alpha})$
for $|\psi_\alpha^{V_0})$.

The optimized effective potential $V_0(\mathbf{r})$ is then found by
variation $\delta E_{\rm SIC} / \delta V_0(\mathbf{r})=0$. 
One can show that the thus optimized $V_0$ can be written as a
sum of three contributions~:
\begin{subequations}
\label{eq:SIC-OEP-double}
\begin{eqnarray}
V_0&=& V_{\rm S} + V_{\rm K} + V_{\rm C} \quad,
\label{eq:V_0}
\\
  V_{\rm S} 
  &=& 
  \sum_i \frac{|{\tilde \varphi_i}|^2}{\rho}v_i 
\quad,
\label{eq:V_S}
\\
V_{\rm K} &=& \sum_i \frac{|{\tilde \varphi_i}|^2}{\rho}({\tilde
  \varphi_i}|V_0-v_i|{\tilde \varphi_i}) 
\quad,
\label{eq:V_K}
\\
V_{\rm C} &=& \frac{1}{2}\sum_i
\frac{\mathbf{\nabla}.(p_i \mathbf{\nabla}|{\tilde
    \varphi_i}|^2)}{\rho} \quad, 
\label{eq:V_C}
\end{eqnarray}
\end{subequations}
with the density $\rho=\sum_{i}|{\tilde \varphi_{i}}|^2$.
The term $V_{\rm S}$ is the Slater contribution, $V_{\rm K} + V_{\rm
S}$ constitutes the KLI approximation, and $V_{\rm C}$ is the
remaining genuine OEP term. {The quantities $v_i$ and $p_i$ entering
  the various contributions are defined as
\begin{subequations}
\begin{eqnarray}
  &&v_i 
  = 
  \sum_\alpha u_{i\alpha}^*{\frac{\tilde{\psi_\alpha}}
  {\tilde{\varphi_i}}} U_\alpha \ ,
\label{eq:vi}
\\
\displaystyle
  &&p_i(\mathbf{r})
  =
  \int\! \textrm d\mathbf{r'}
  \{V_0(\mathbf{r'})\!-\!v_i(\mathbf{r'})\}
  \frac{{\tilde \varphi_i}(\mathbf{r'})}{{\tilde \varphi_i}(\mathbf{r})} 
  G_i(\mathbf{r},\mathbf{r'})
  \ ,
\label{eq:pi}
\\
\displaystyle
  &&G_i(\mathbf{r},\mathbf{r'})
  =
  \sum_{j\ne i}
  \frac{{\tilde \varphi_j^*}(\mathbf{r}){\tilde \varphi_j}(\mathbf{r'})}{\varepsilon_j-\varepsilon_i}
  \ .
\label{eq:pi2}
\end{eqnarray}
\end{subequations}
}
Inserting (\ref{eq:unitrans}) and (\ref{eq:vi}) in the quantities
involved in (\ref{eq:V_S}-\ref{eq:V_C}), one obtains~: 
\begin{subequations}
\begin{eqnarray}
&&  V_{\rm S} = \sum_\alpha \frac{|{\tilde \psi_\alpha}|^2}{\rho}U_\alpha
  \quad,
  \label{eq:V_Sbis}\\
&&
  ({\tilde \varphi_i}|V_0-v_i|{\tilde \varphi_i}) 
  = 
  \sum_{\alpha,\beta}u_{i\alpha}^*u_{i\beta}
  ({\tilde \psi_\beta}|V_0-U_\alpha|{\tilde \psi_\alpha})
  \,,
  \label{eq:inV_K}\\
&&
  p_i(\mathbf{r})
  =
  \sum_\alpha\!u_{i\alpha}^*\int\!\!\textrm d\mathbf{r'} 
  (V_0(\mathbf{r'})\!-\!U_\alpha(\mathbf{r'}))
\nonumber\\[-3pt]
&&\hspace*{10em}
  {\tilde \psi_\alpha}(\mathbf{r'})G_i(\mathbf{r},\mathbf{r'})
  \quad,
  \label{eq:p_i}
\end{eqnarray}
\end{subequations}

We now exploit the interesting property that the 
${\tilde \psi_\alpha}$ are localized \cite{Ped84}.
This means that, at a given $\mathbf r$, one single
${\tilde \psi_\alpha}$ mostly dominates the other wavefunctions
${\tilde \psi_\beta}$. This amounts to have almost vanishing (\ref{eq:inV_K})
and (\ref{eq:p_i}), and thus allows to neglect in $V_0$ the two terms
$V_{\rm K}$ and $V_{\rm C}$ defined in Eqs.~(\ref{eq:V_K}) and
(\ref{eq:V_C}) respectively. 
The final OEP result thus naturally reduces to
\begin{subequations}
\label{eq:GenSlat}
\begin{eqnarray}
  \varepsilon_i {\tilde \varphi_i}({\bf r})
  &=&
  \left[\hat{h}^\mathrm{(LDA)}({\bf r})-V_0(\mathbf{r})\right] {\tilde \varphi_i}({\bf r})  
  \quad,
\label{eq:mfOEP}
\\
  V_0
  &\simeq&
  \sum_\alpha \frac{|{\tilde \psi_\alpha}|^2}{\rho}U_{\rm LDA}[|{\tilde \psi_\alpha}|^2]
  \quad,
\label{eq:Slater3}
\\
  0
  &=&
 ({\tilde \psi_{\beta}}|U_{\rm LDA}[|{\tilde \psi_\alpha}|^2]
               -U_{\rm LDA}[|{\tilde \psi_\beta}|^2]|{\tilde \psi_\alpha}) 
\,,
\label{eq:symcond3}
\\
  {\tilde \psi_\alpha}
  &=&
  \sum_i {\tilde \varphi_i} \, u_{i\alpha}
\quad,
\label{eq:unitra}
\end{eqnarray}
\end{subequations}
with the symmetry condition (\ref{eq:symcond3}) explicitely rewritten
for completeness.
The mean field equation (\ref{eq:mfOEP}) generates the set ${\tilde
\varphi_i}$ of occupied states, while the unitary transformation
(\ref{eq:unitra}) serves to accommodate the symmetry condition
(\ref{eq:symcond3}) which, in turn, defines the ``localized'' set
${\tilde \psi_\alpha}$ that enters the Slater mean field $V_0$ as
given in Eq.~(\ref{eq:Slater3}).
Note that this equation has the form of a Slater
approximation \cite{Kri90a,Slat51}
but is constructed from the localized ${\tilde \psi_\alpha}$ and applied 
to the ${\tilde \varphi_i}$. We thus call this new scheme 
\emph{``Generalized SIC Slater''} approximation, which differs from
the usual Slater scheme because of the two basis sets involved here.

\subsection{Recovering "standard" SIC-OEP}

The SIC-OEP scheme presented in section \ref{sec:sic_oep_2sets}
differs from usual SIC-OEP formulations in that it employs
simultaneously two complementing sets of orbitals ${\tilde
\psi_\alpha}$ and ${\tilde \varphi_i}$. 
It is instructive to
step back to "standard" SIC-OEP. Indeed the recovery procedure 
sheds some light on the relevance of approximations performed
in standard SIC-OEP approaches. 
One recovers the standard formulation by simply setting the unitary transform 
Eq.~(\ref{eq:unitrans}) to unity, i.e.,
\begin{equation}
u_{i\alpha} \rightarrow \delta_{i\alpha}
\label{eq:ut221}
\end{equation}
which renders the sets ${\tilde \psi_\alpha}$ and ${\tilde \varphi_i}$
identical.
The symmetry condition becomes obsolete because there is no unitary
transformation to be optimized.
The crucial $v_i$ then simply reduces to $U_{\rm LDA}[|{\tilde
\varphi_i}|^2]$, see Eq.~(\ref{eq:SICmf}), and the usual Slater and
KLI approximations are 
directly obtained from Eqs.~(\ref{eq:V_0}-\ref{eq:V_K}).
KLI means to approximate $V_0\approx V_\mathrm{S}+V_\mathrm{K}$
while the usual Slater approximation goes one step further to
$V_0\approx V_\mathrm{S}$, but then applied in a situation with
only one set of wavefunctions.

The usual reasoning to validate the KLI approximation, i.e.,
neglecting the correction term~(\ref{eq:V_C}), relies on the
oscillating nature of ${\bf p}_i({\bf r})$ which makes it small in the
average. Going from KLI to Slater is then usually justified only in
the case of homogeneous or well localized systems {(described
by ${\tilde \varphi_i}$)}.  Our Generalized Slater potential, as built
from localized orbitals {(described by ${\tilde
\psi_\alpha}$)}, thus naturally appears as a good and well justified
approximation to full OEP.  It remains to see how it performs in
actual computations.

\section{Test for a 1D dimer molecule}

As first applications of the generalized Slater approximation,
we consider two test cases. First, we consider
a dimer molecule in one dimension (1D) which is a
widely used schematic model for critically probing
crucial structural and dynamical
features of an approach, see e.g.  \cite{Hen98}.  
In particular the studies of
\cite{Mun06} have shown that 1D models carry all features of
(de-)localization. The two electrons in the dimer have the same spin
in order to test the exchange term and the self-energy error. As an
interaction, we use a smoothed Coulomb potential,
$ \displaystyle w_{\rm ij} = {e^2}/{\sqrt{(x_{\rm i}-x_{\rm
j})^2 + a_{\rm ij}^2}} 
$, 
where the parameters $a_{\rm ij}$ for electron-electron, electron-ion
and ion-ion interactions are tuned to reproduce typical molecular
energies. Taking this interaction, we develop within LDA an energy
functional for the exchange term. Working at the level of exchange
only allows to have Hartree-Fock (HF) calculations as a benchmark to which
approximations can be compared. 
The second test case will be naturally a "localized" system
with more electrons, namely a C atom with 4 electrons 
(3 spins up and 1 down),  computed in realistic three
dimensions.

\begin{figure}[htbp]
\begin{center}
\epsfig{file=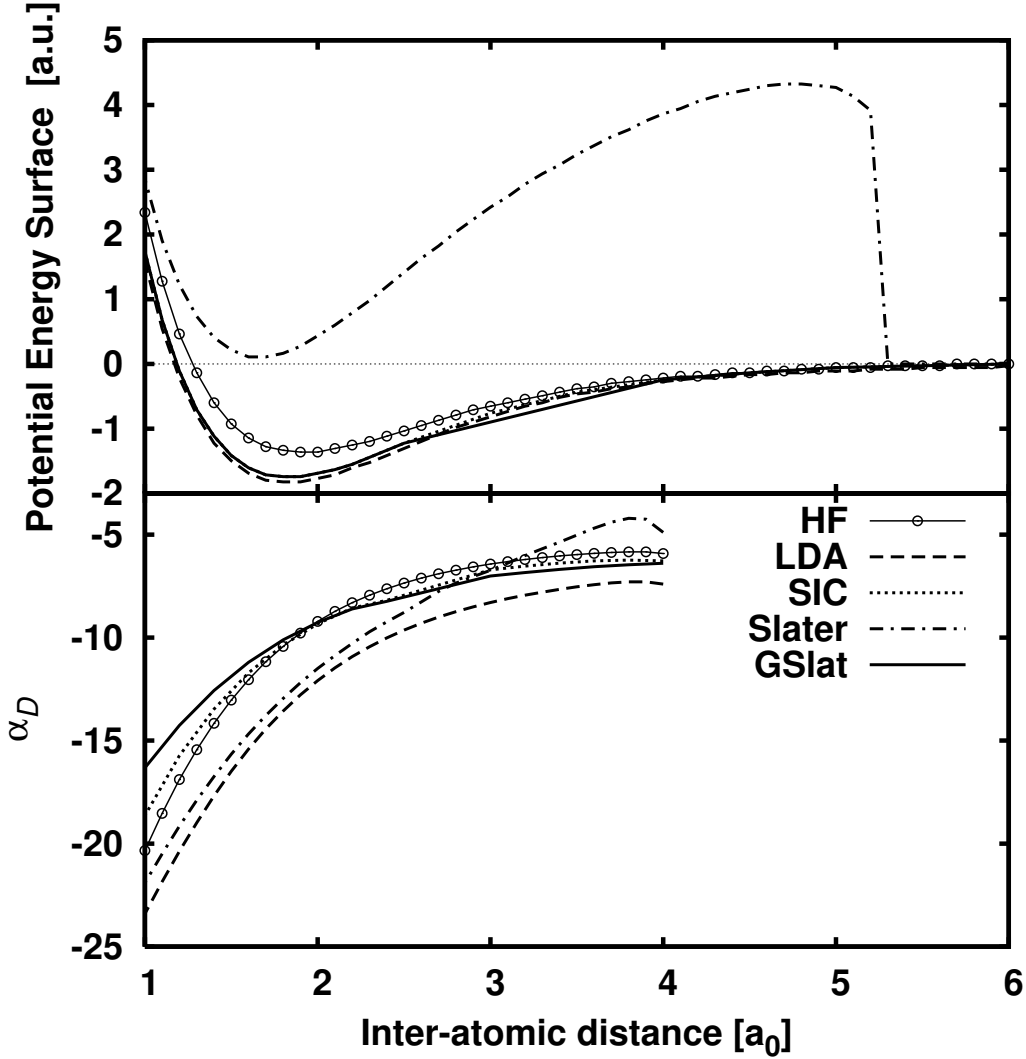,width=\linewidth}
\caption{\label{fig:pes-pol} 
Upper~: Born-Oppenheimer surface for a 1D dimer molecule with two electrons.
Compared are results from HF, LDA, SIC, usual Slater approximation,
and the generalized Slater approximation (GSlat).
Lower~: Polarization as a function of inter-atomic distance for
the five approximations considered.
}
\end{center}
\end{figure}
Fig. \ref{fig:pes-pol} shows the result for the dimer case.
One finds in the upper panel the dimer binding energy
$E_\mathrm{bond}=E_\mathrm{dimer}-2E_\mathrm{atom}$ as a function of
atomic distance for the various approximations.  The lower panel shows the
dipole polarizability $\alpha_D$ which is a sensitive test for 
DFT considerations \cite{Gri01a,Gru02a}.
The dimer binding energy in Slater approximation (dashed-dotted curve)
shows an unnatural bump with a maximum at distance of 5 $a_0$. This is
caused by a trend to delocalization in this approximation. The same
feature happens in KLI (not shown here). SIC instead, which has a
tendency to localization, produces a reasonable energy curve (dotted
line under the full curve) close to the exact HF case (full line). It
is gratifying to see that the generalized Slater approximation (full
curve labeled ``GSlat'', almost identical to the SIC one) performs
equally well. The simple-most LDA does also perform fairly well what
the dimer binding curve is concerned (dashed line almost identical to
the SIC one).
A different perspective is provided by the results for the
polarizability. LDA is notoriously off the benchmark HF result.  SIC
performs very well. SIC-Slater, however, fails again.  But generalized
Slater (full curve) does follow the SIC result nicely for 
physically relevant bond distances.

\begin{figure}[htbp]
\begin{center}
\epsfig{file=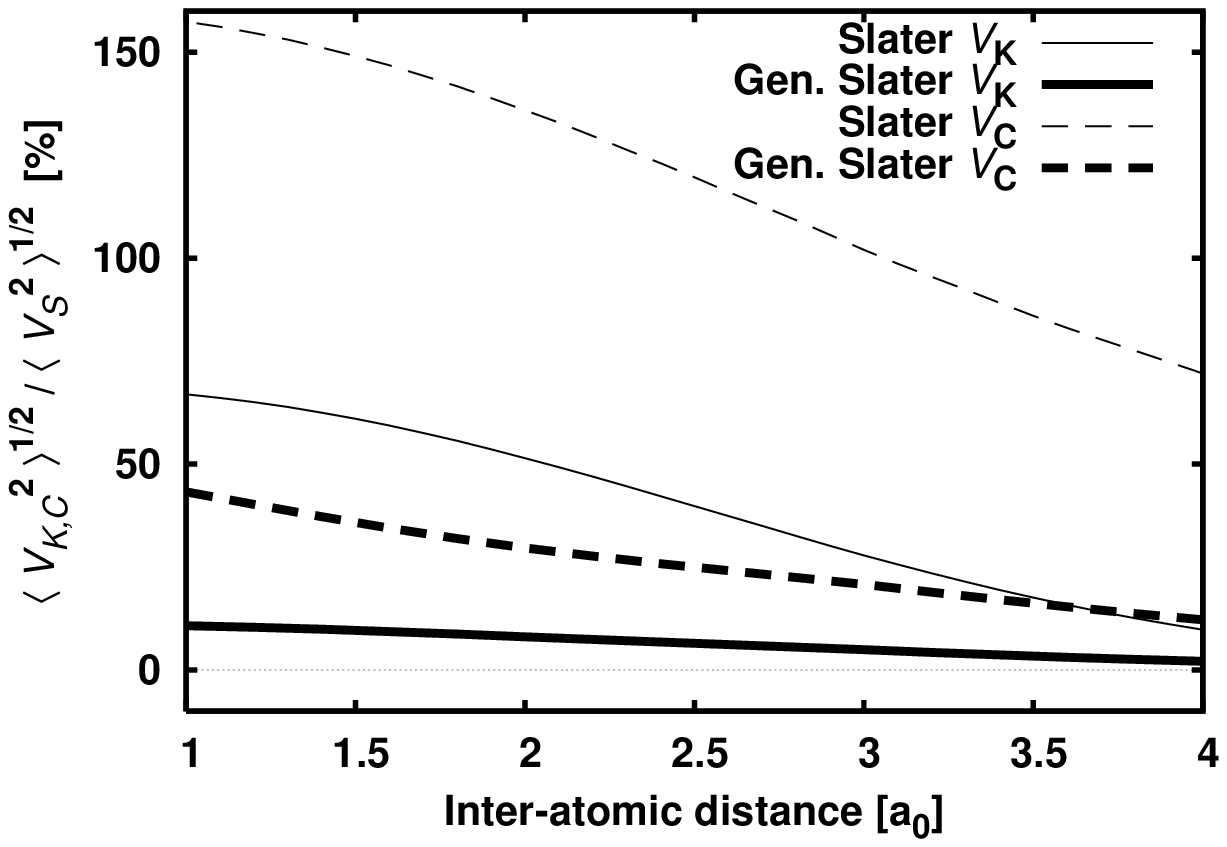,width=\linewidth}
\caption{\label{fig:oep_terms} 
\MDcomm{R.m.s.} \PGRcomm{expectation value}
of the neglected KLI-OEP terms, $V_{\rm K}$ and $V_{\rm
  C}$, rescaled by that of the Slater potential $V_{\rm S}$~:
in the usual Slater approximation (thin curves), see
Eqs.~(\ref{eq:V_K},\ref{eq:V_C}), and in our generalized Slater one
(thick curves), calculated as in Eqs.~(\ref{eq:V_K},\ref{eq:V_C}) but
with the $\tilde \psi_\alpha$ instead of the ${\tilde \varphi_i}$.
Results for the 1D dimer molecule with two electrons are drawn
versus inter-atomic distance and are given in percentages.
}
\end{center}
\end{figure}
Fig.~\ref{fig:oep_terms} analyzes the findings by comparing 
the terms neglected when stepping from OEP down to (generalized)
Slater. \MDcomm{\JMcomm{Mainly} because of the oscillating behavior of $p_i$ defined
in Eq.~(\ref{eq:pi}), the mean values of $V_{\rm K}$ and $V_{\rm C}$
are 
\JMcomm{very small}. Thus we plot in Fig.~\ref{fig:oep_terms} their
root-mean-square \PGRcomm{expectation value, i.e.
$\sqrt{\langle V_i^2\rangle}$
with $\langle V_i^2\rangle=\int \textrm dx\, V_i^2(x)$,
}, normalized by that of $V_{\rm S}$.}
\MDcomm{The r.m.s. }\PGRcomm{expectation value of} the KLI term $V_\mathrm{K}$ is generally
smaller than \MDcomm{that of} $V_\mathrm{C}$. The gain in smallness by
generalized Slater is 
dramatic (mind the percentage scale). Even there, one 
observes some growth of \MDcomm{the r.m.s.} 
\PGRcomm{expectation value} of $V_\mathrm{C}$ for
small distances. This coincides with the deviation in the
polarizability observed in Fig.~\ref{fig:pes-pol} for very small
distances.

\begin{figure}[htbp]
\begin{center}
\epsfig{file=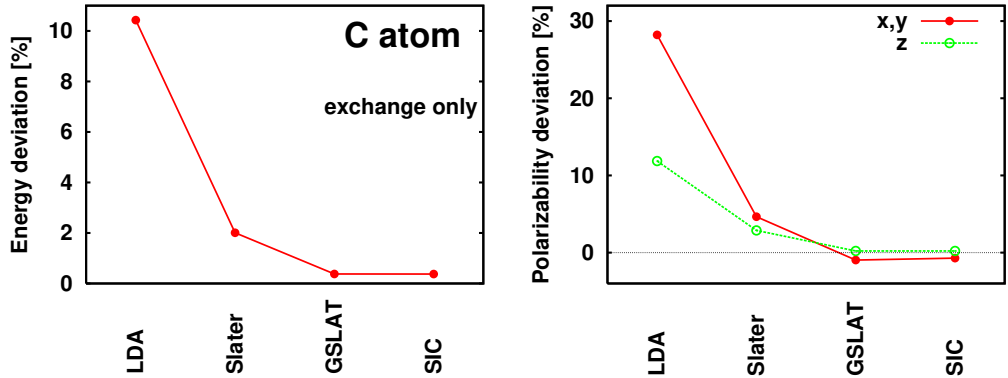,width=\linewidth}
\caption{\label{fig:C-atom-relative} 
Comparison of various approaches for  the C atom.
Left panel: Deviation of binding energy from the
Hartree-Fock benchmark value.
Right panel: Deviation of dipole polarizabilities
(in $x$, $y$, and $z$ direction)
from the Hartree-Fock values. 
}
\end{center}
\end{figure}
As a test case for a compact electron distribution, we have considered
the C atom using fully three dimensional calculations \cite{Cal00}. 
The starting point
is a calculation with exact exchange and the approximations are
compared consistently with LDA for exchange only. 
Fig.~\ref{fig:C-atom-relative} shows the relative deviations for
energy and polarizabilities $\alpha$ (note that the electron cloud is
slightly asymmetric such that $\alpha$ depends on the spatial
direction). The conclusion is obvious. SIC is a good approximation to the
exact calculation and generalized Slater does equally well, while
simple Slater or LDA show larger deviations.

\section{Conclusion and outlook}

We have presented a formulation of SIC-OEP which employs two different
sets of $N$ single-particle wavefunctions. One set is taken for the
solution of the OEP equations, thus diagonal in energy and most likely
delocalized. The other set is used in setting up the SIC energy which
becomes minimal for localized wavefunctions. Both sets are
connected by a unitary transformation which leaves key features as,
e.g, the total density invariant. Using that double set allows to
accommodate two conflicting demands, energy diagonality versus
locality. The unitary transformation is determined by minimization of
the SIC energy which leads to what we called the symmetry condition, a
key building block of the SIC equations.
The localized character of the SIC-optimizing set is well suited to
justify the steps from OEP to KLI and further to Slater
approximation. We call that scheme generalized Slater approximation.
By virtue of the double-set technique, it has a wider range of
applicability than straightforward KLI or Slater approximation.
We have tested the scheme on a 1D model of a dimer at various
distances and on the C atom in full 3D, comparing LDA, standard Slater
approximation, generalized Slater approximation, and SIC with full
Hartree-Fock as a benchmark.  The numerical tests demonstrate that the
generalized Slater approximation performs remarkably well.
The present development is to be considered as a first step.  A
generalization to a time-dependent version is obvious, but raising
several new consistency conditions (energy conservation, stability).
Work in that direction is in progress.
 
This work was supported,
by  
%
Agence Nationale de la Recherche (ANR-06-BLAN-0319-02), 
the Deutsche Forschungsgemeinschaft (RE 322/10-1),
and the  Humboldt foundation.

\end{document}